\documentclass{edp-conf}
\usepackage{graphicx}
\begin{document}

\TitreGlobal{SF2A 2003}

\title{The accuracy of the ALI method for solving the radiative transfer equation in stellar atmospheres}
\author{Chevallier, L.}\address{Centre de Recherche Astronomique de Lyon (UMR
5574 du CNRS), Observatoire de Lyon, 9, avenue Charles Andr\'{e},
69561 Saint-Genis-Laval Cedex, France.\\
Email: \texttt{loic.chevallier@obs.univ-lyon1.fr}, \texttt{rutily@obs.univ-lyon1.fr}}
\author{Paletou, F.}\address{Observatoire Midi-Pyr\'{e}n\'{e}es, Laboratoire d'Astrophysique (UMR 5572 du CNRS), 14 avenue E. Belin - 31400 Toulouse Cedex, France. Email: \texttt{fpaletou@ast.obs-mip.fr}}
\author{Rutily, B.$^1$}
\runningtitle{The accuracy of the ALI method}
\setcounter{page}{1}
\index{Chevallier, L.}
\index{Paletou, F.}
\index{Rutily, B.}

\maketitle
\begin{abstract}
We test the accuracy of the ALI method, widely used in stellar atmosphere modelling, by solving exactly a standard radiative 
transfer problem in plane-parallel geometry. Some recommendations are given for a practical use of this method.
\end{abstract}
%
\section{Introduction}
We are interested in solving the radiative transfer equation in a homogeneous and isothermal plane-parallel atmosphere 
in local thermodynamical equilibrium, assuming isotropic and monochromatic scattering. If no radiation is incident on 
the boundary layers, the thermal source function is $\epsilon B(T)$, where $\epsilon$ is the photon destruction probability 
per scattering and $B(T)$ is the Planck function at temperature $T$. The source function $S$, normalized to the Planck function, 
satisfies the following integral equation
\begin{equation}
\label{eq_source}
S(\tau)=\epsilon + (1-\epsilon) (\Lambda S)(\tau)\,,
\end{equation}
where
\begin{equation}
\label{eq_lambda}
(\Lambda S)(\tau)=\frac{1}{2}\int_0^{\tau^*}E_1(\vert \tau-\tau^{\prime}\vert)S(\tau^{\prime}) \,\mathrm{d}\tau^{\prime}\, .
\end{equation}
$E_1(\tau)=\int_0^1\exp(-\tau/\mu)\,\mathrm{d}\mu /\mu$ is the first exponential integral function, and the 
optical depth variable $\tau$ covers the range $[0, \tau^*]$, $\tau^*$ being the optical thickness of the atmosphere at 
a given frequency. Equation (\ref{eq_source}) models the multiple scattering of photons in a continuum or in a 
spectral line with a rectangular profile.

Chevallier and Rutily (2003) have solved recently Eq. (\ref{eq_source}) with an accuracy better than $10^{-10}$ for any 
value of $\epsilon$, $\tau^*$ and $\tau$. Their solution can safely be used as a reference solution for testing the 
accuracy of a numerical code, e.g., the ALI code widely used in stellar atmosphere modelling. Our focus is to present some 
tests for this code, which we first describe in the next section. Some additional tests can be found in Chevallier et al. (2003). 
\section{Brief description of our ALI code}
The ALI code used in this paper is a combination of an accelerated iterative method (with a diagonal $\Lambda$-operator) and a formal solver based on parabolic short characteristics.
Recent reviews on this approach are Paletou (2001) and Hubeny (2003).
The grid we chose for the integration with respect to $\tau$ over $[0,\tau^*]$ is logarithmic and symmetric with respect to the mid-plane, with $n_\tau=9$ points per decade starting from $\tau=10^{-4}$ and including the $\tau=0$ point.
The $\Lambda$-operator is defined by an integration over a directional variable $\mu$ in $[-1,1]$, which is performed with the help of a Gaussian quadrature ($n_\mu=5$ Gauss-Legendre points on $[0,1]$, plus opposite points).
The iterations can be stopped after $N \geq N_\mathrm{c}$ iterations, where $N_\mathrm{c}$ is the number of iterations required for the accuracy $d_\mathrm{M}$ (i.e., maximal relative error) to reach the convergence threshold (Fig. 1).
As $N_\mathrm{c}$ is not known \textit{a priori}, the iterations are usually stopped when the relative change $R_\mathrm{c}$ of the source function between two successive iterations is less than a given small number, for instance $10^{-3}$.
\begin{figure}[h]
\label{fig_convergence}
\centering
\includegraphics[width=9cm]{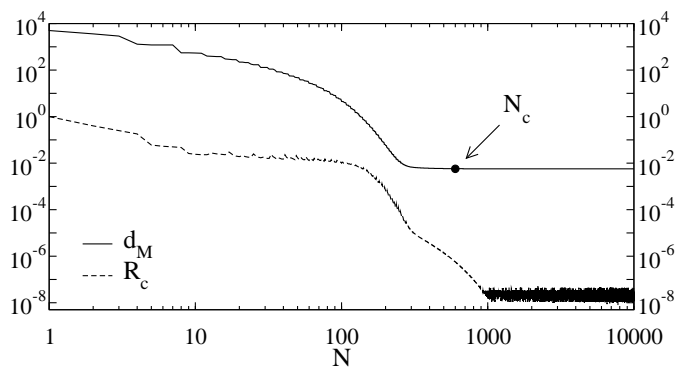}
\caption{Accuracy $d_\mathrm{M}$ and relative change $R_\mathrm{c}$ of the ALI code as a function of the number of iterations $N$ for a ``strong'' spectral line $\epsilon =10^{-8}$, $\tau^*=2\times 10^8$.} 
\end{figure}
\section{The accuracy tests} 
Figure 2 shows the relative error of our ALI code as a function of $\tau$ for $N=1000 > N_\mathrm{c}$.
The three curves correspond to typical values of $(\epsilon, \tau^*)$ in a continuum $(\epsilon = 0.01, \tau^* = 20)$, in an ``average'' spectral line $(\epsilon = 10^{-4},\tau^* = 2000)$ and in a ``strong'' spectral line $(\epsilon = 10^{-8}, \tau^* = 2\times10^8)$.
The relative error is maximal when $\tau$ is smaller than the \textit{thermalization depth} $1/k(\epsilon)$, equal to 6, 58 and 5774 for the continuum, average line and strong line respectively (Chevallier and Rutily 2003).
It significantly improves in the deep layers of the star, up to $10^{-8}$.
We confirm that the accuracy of our ALI code for $N > N_\mathrm{c}$ does not depend on $\epsilon$ and $\tau^*$, but only on the refinement of the $\tau$-grid and the $\mu$-grid.

Figure 3 shows the number of iterations $N_\mathrm{c}$ needed to achieve convergence, for an extended range of $(\epsilon,\tau^*)$. 
We note an important slowing down of the convergence when $\epsilon\to 0$ and $\tau^*\to\infty$.
From preceding studies of the semi-infinite case, it is well known that $N_\mathrm{c}$ strongly depends on $\epsilon$ (Trujillo Bueno \& Fabiani Bendicho 1995).
The present study of the finite case clarifies the influence of $\tau^*$: $N_\mathrm{c}$ depends only on $\epsilon$ when $k(\epsilon)\tau^* > 100$ and drops down for weaker values of $\tau^*$.
\begin{figure}[h]
\label{fig_relerr_tau}
\centering
\includegraphics[width=9cm]{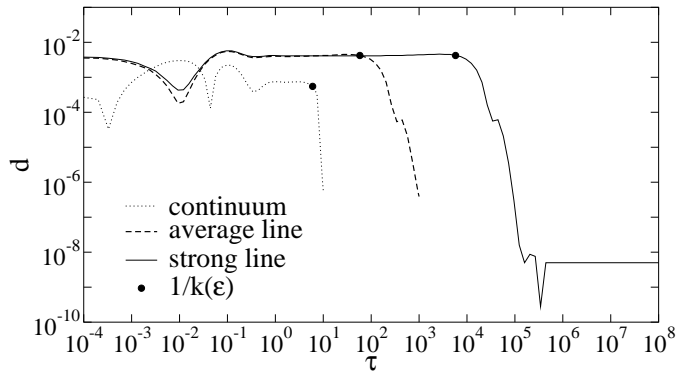}
\caption{Relative error $d(\tau)$ of the ALI code for $N=1000>N_\mathrm{c}$. Black dots show the thermalization depth $1/k(\epsilon)$ for each case ($k(\epsilon)\sim \sqrt{3\epsilon}$ when $\epsilon\to 0$).}
\end{figure}
\begin{figure}[h]
\label{fig_nc2d}
\centering
\includegraphics[width=9cm]{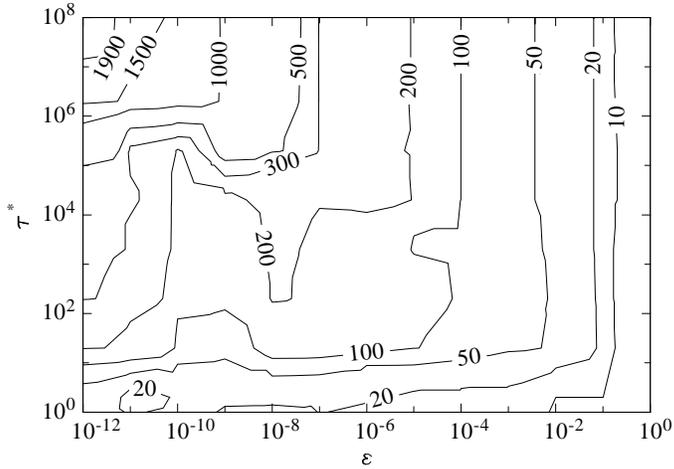}
\caption{Number of iterations $N_\mathrm{c}$ required for convergence as a function of $\epsilon$ and $\tau^*$.}
\end{figure}

In practice, an accuracy better than 1\% is generally required while choosing  $R_\mathrm{c} = 10^{-3}$ as the iteration stopping criterion.
Figure 4 shows the accuracy of ALI for a set of slabs such as $\epsilon=10^{-p}$ and $\tau=2\times 10^p$ ($p=1,\ldots, 8$) when $R_\mathrm{c} = 10^{-3}$ and $R_\mathrm{c} = 10^{-4}$.
The criterion $R_\mathrm{c} = 10^{-3}$ does not warrant an accuracy of 1\% for any $p$.
We thus recommend the value $R_\mathrm{c} \le 10^{-4}$ to ensure an accuracy of 1\% in stellar atmosphere calculations.
\begin{figure}[h]
\label{fig_rcpractical}
\centering
\includegraphics[width=9cm]{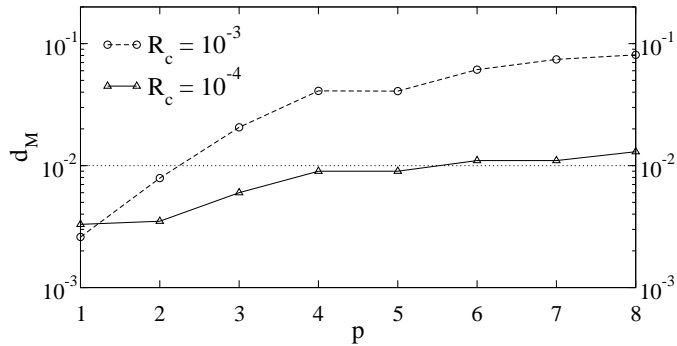}
\caption{Accuracy $d_\mathrm{M}$ as a function of parameter $p$, where $\epsilon=10^{-p}$ and $\tau^*=2\times 10^p$, for $R_c=10^{-3}$ and $R_c=10^{-4}$. The dotted-line figures out the 1\% accuracy level.}
\end{figure}
\section{Conclusion}
In this particular problem (\ref{eq_source}), our code was able to converge in all situations, its accuracy being bounded 
by $5\times 10^{-3}$ and $10^{-2}$ when using the stopping criterion $R_c = 10^{-4}$. We intend to extend this work to more 
realistic problems (e.g., non isothermal atmospheres), replacing the Jacobi scheme by Gauss-Seidel or Successive Over-Relaxation schemes (see Trujillo Bueno \& Fabiani Bendicho 1995).

\end{document}